\documentclass[prb,twocolumn,showpacs]{revtex4}

\usepackage{graphicx}
\usepackage{subfigure}
\usepackage{amsfonts}
\usepackage{amsmath}
\usepackage{amssymb}

\usepackage{color}

\usepackage{hyperref}

\newcommand{\HA}{{Anderson-Hubbard }}

\begin{document}

\title{Self-consistent study of Anderson localization in the \HA model in two and three dimensions}

\author{Peter \surname{Henseler}}
\author{Johann \surname{Kroha}}
\affiliation{Physikalisches Institut, Universit\"at Bonn, Nussallee 12, 53115 Bonn, Germany}
\author{Boris \surname{Shapiro}}
\affiliation{Department of Physics, Technion-Israel Institute of Technology, Haifa 32000, Israel}

\begin{abstract}
We consider the change in electron localization due to the presence of electron-electron repulsion in the \HA model. Taking into account local Mott-Hubbard
physics and static screening of the disorder potential, the system is mapped 
onto an effective single-particle Anderson model, which is studied within 
the self-consistent theory of electron localization. We find rich
nonmonotonic behavior of the localization length $\xi$ in two-dimensional 
systems, including an interaction-induced exponential enhancement of  
$\xi$ for small and intermediate disorders although $\xi$ remains finite.
In three dimensions we identify for half filling a Mott-Hubbard-assisted  
Anderson localized phase existing between the metallic and the
Mott-Hubbard-gapped phases. For small $U$ there is re-entrant behavior
from the Anderson localized phase to the metallic phase.  
\end{abstract}

\pacs{71.10.Fd, 73.20.Fz, 72.15.Rn, 71.30.+h} 

\maketitle

\section{Introduction}
The research on interacting, disordered systems is one of the central topics in
today's condensed-matter physics. In particular, the experimental signatures
of a two-dimensional (2D) metal-insulator transition in dilute disordered
electron systems,\cite{ks04} absent in the single-particle theory of 
Anderson localization,\cite{and58, aalr79} have triggered
many theoretical research activities. As a prototype model for the interplay
of strong correlations and randomness, the \HA model has been
intensely studied.\cite{sbs03}$^{-}$\cite{kns08}
While in electron systems the bare Coulomb interaction is long ranged,
inducing the additional question of the range of the effective interaction 
in a disordered system, the \HA model assumes an on-site interaction
which may grasp many of the salient features of the electron-electron interaction. 
Disordered many-body systems with tailor-cut short-range interactions have
most recently been realized experimentally as cold atomic gases in random
potentials.\cite{bec08}  
Despite these efforts, even in systems with short-range interaction
the existence of a metallic ground state in $d\leq2$ dimensions has remained elusive. Recent numerical works showed that the presence of interactions can at least suppress significantly the localizing effect of the disorder.\cite{sbs03}$^{-}$\cite{her01,np05}

One of the ideas proposed to explain the delocalizing effect is screening of
the random potential by the interaction, discussed controversially in the
literature. A detailed analysis of this screening effect is the subject of this paper. 
For that purpose, our approach starts from the atomic limit of the \HA model. 
In the presence of a Hubbard interaction $U$, the on-site energy of a particle
depends on whether this site is occupied by another particle or not, leading
to an interaction-induced renormalization of the distribution of on-site  
energy levels. Below we argue that this effective disorder distribution of 
the atomic limit will still provide a good description of the static 
interaction effects when a finite hopping amplitude is 
included in the full \HA model as long as the 
fluctuations of the on-site energy levels are large compared
to the kinetic band energy (hopping amplitude). 
In particular, this will hold for arbitrarily large Hubbard interaction 
even when the localization length becomes large. As seen below, the latter
may occur even for relatively strong disorder near the Anderson transition
in three dimensions and as an interaction effect in two dimensions. 
By this reasoning, the \HA model is reduced to an effective single-particle 
Anderson disorder model\cite{and58} with renormalized level distribution, 
for which we calculate ensemble-averaged single-particle and transport 
properties at temperature $T=0$. Clearly, inelastic as well as virtual 
interaction effects are neglected by this approach. 
For the present purpose they may, however, be unimportant due 
to the vanishing quasiparticle relaxation rate\cite{aa85} at the Fermi energy 
in 2D and three-dimensional (3D) disordered systems for $T=0$.

In a previous work,\cite{hks08} we already presented an analytical study of
this approach, exploiting an exact relation\cite{tho72} between the 
localization length and the ensemble-averaged single-particle density of states (DOS) 
in one dimension.
In accordance with numerical results obtained for the \HA model,\cite{sbs03}$^{-}$\cite{swa08} we
could demonstrate that weak interaction reduces the effective disorder
(screening) while a strong interaction effectively enhances the localization,
corresponding to a hopping suppression by interaction (Mott-Hubbard physics). 

In this paper, we extend our analysis by
applying the self-consistent theory of Anderson 
localization,\cite{vw80}$^{-}$\cite{kop84} 
which allows for a quantitative analysis of the effective single-particle 
Anderson disorder model in a broad parameter regime, particularly in two and 
three dimensions. Despite the simplifications made in our approach, we find good 
agreement with recent numerical studies, especially in the Anderson localized 
regime of one and two dimensions. Our analytical approach therefore allows for
a critical assessment of some of the conclusions drawn from the numerical 
results. Our results indicate that screening of the disorder 
seems to be the most relevant physical mechanism for the interaction-induced 
delocalization effect in the disorder localized regime of the \HA model.

\section{Atomic-Limit Approximation}\label{sec al}
We consider the \HA model for fermions at zero temperature on a hypercubic lattice in $d$ dimensions with lattice spacing $a$. It is described by the Hamiltonian
\begin{eqnarray}\label{HA_Hamiltonian}
 H & = & H_{0} \: + \: H_{\textrm{kin}} \: + \: H_{\textrm{e-e}} \nonumber \\ 
 & = & \sum\limits_{i,\sigma} \left(\varepsilon_{i}^{\phantom{\dagger}}-\mu\right) c_{i\sigma}^{\dagger} c_{i\sigma}^{\phantom{\dagger}} \: - \: t\!\sum\limits_{<i,j>,\sigma} \!c_{i,\sigma}^{\dagger} c_{j\sigma}^{\phantom{\dagger}} \nonumber \\
 & & \: + \: U \sum\limits_{i} n_{i\uparrow}^{\phantom{\dagger}} n_{i\downarrow}^{\phantom{\dagger}} \, .	
\end{eqnarray}
$c_{i\sigma}^{\dagger} (c_{i\sigma}^{\phantom{\dagger}})$ are
creation (destruction) operators of a fermion at site $i$ with spin
$\sigma$ and $n_{i\sigma}^{\phantom{\dagger}}=c_{i\sigma}^{\dagger}
c_{i\sigma}^{\phantom{\dagger}}$. $t$ is the nearest-neighbor hopping
amplitude, $U>0$ is the on-site repulsion, and $\mu$ is the chemical potential. The
on-site energies $\{\varepsilon_{i}^{\phantom{\dagger}}\}$ are assumed to be
independent random variables with a box probability distribution 
$p(\varepsilon) = \Theta(\Delta/2-|\varepsilon|)/\Delta$, where 
the disorder strength is parametrized by the width $\Delta$.
The lattice filling, i.e., the average particle number per lattice site (summed over spin), 
will be denoted by $\rho$. \\

\begin{figure}
 \begin{center}
  \includegraphics[scale=0.2]{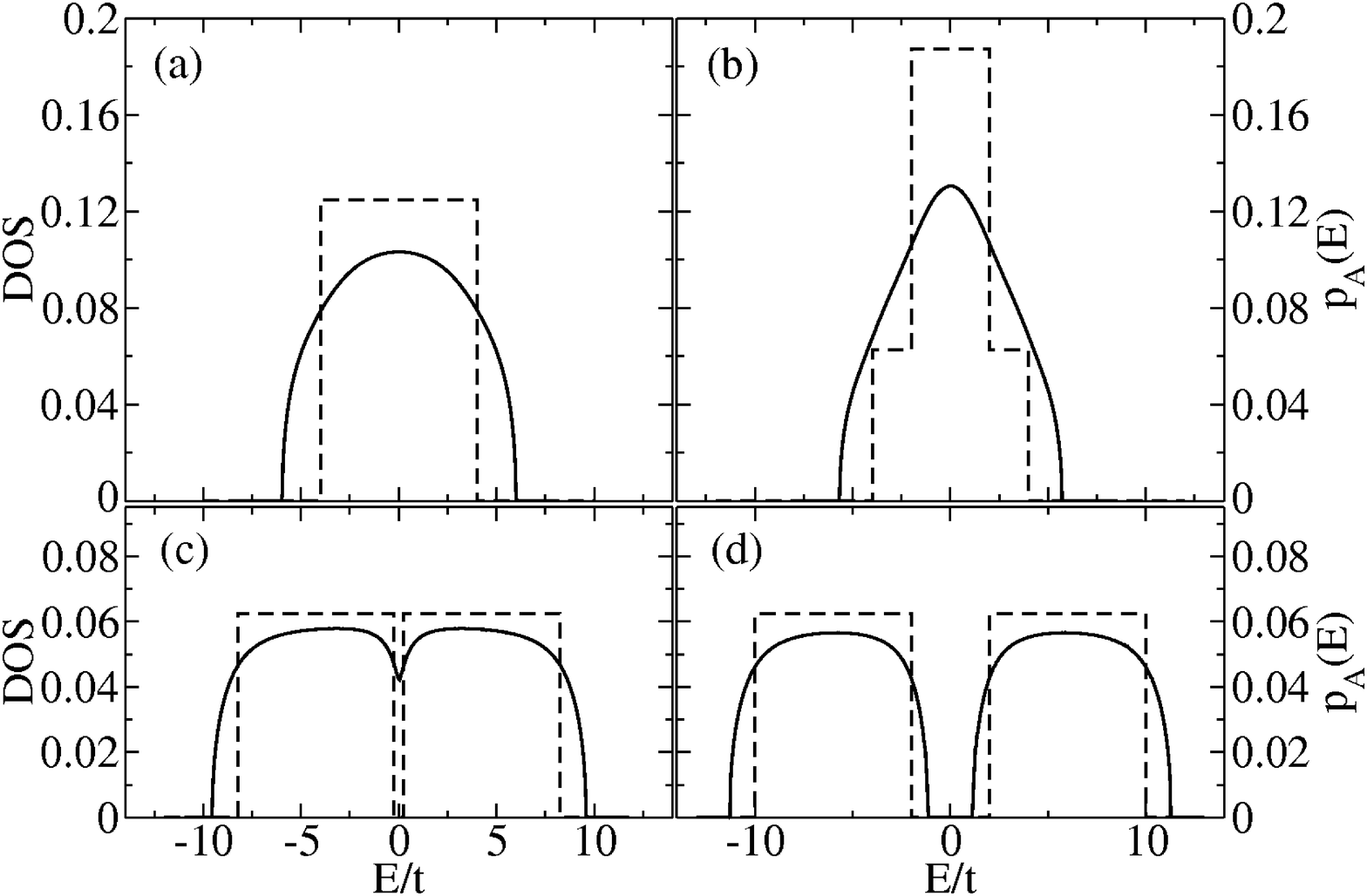}
  \caption{CPA density of states (solid lines) and renormalized site-energy distribution $p_{A}(\varepsilon)$ (dotted lines) in $d=2$ at half filling for $\Delta/t=8$ 
and (a) $U/t=0$, (b) $4$, (c) $8.5$, and (d) $12$. Energies are measured relative to the Fermi level.}
\label{fig dos}
 \end{center}
\end{figure}

In the atomic limit, $t=0$, the energy of a particle on site $i$ depends on
the occupation number $n_i$ of that site and can be obtained in the 
paramagnetic phase by shifting the bare on-site energy $\varepsilon_i$ 
according to\cite{hks08}
\begin{eqnarray}\label{eqn al distr}
 \varepsilon_{i}^{\phantom{\dagger}} & \mapsto & \!\!\left\{\!\!
 \begin{array}{rcl} \varepsilon_{i}^{\phantom{\dagger}}+U \, &
 \mbox{if} & \varepsilon_{i}^{\phantom{\dagger}} \leq \mu - U \\
 \begin{array}{r} \varepsilon_{i}^{\phantom{\dagger}} + U \\
 \varepsilon_{i}^{\phantom{\dagger}} \end{array} & \mbox{if} & \mu -
 U < \varepsilon_{i}^{\phantom{\dagger}} \leq \mu \: \left( \hspace*{-0.3ex} \begin{array}{l}
 \mbox{each with} \\ \mbox{prob. of $\frac{1}{2}$} \end{array} \hspace*{-0.2ex} \right) \\
 \varepsilon_{i}^{\phantom{\dagger}} \, &  \mbox{if} &
 \varepsilon_{i}^{\phantom{\dagger}} > \mu\ . \end{array} \right. \quad
\end{eqnarray}
The probability distribution $p_{A}(\varepsilon)$ of these renormalized energy
levels is, hence, identical to the (averaged) spectral density of the 
Hubbard two-pole Green's function in the atomic limit (shifted by the chemical potential). 
Examples of this renormalized distribution are given in Fig.~\ref{fig dos} and in Ref.~\onlinecite{hks08}.
There it was shown that for $U\ll\Delta$ the variance of $p_{A}(\varepsilon)$
as compared to $p(\varepsilon)$ is reduced. On the other hand,
it is evident from Eq.~(\ref{eqn al distr}) that for $U\geq \Delta $ 
the support of the distribution $p_{\rm A}(\varepsilon)$ splits into
two disconnected intervals, leading eventually to a Mott-Hubbard gap in the 
averaged DOS of the model [Eq.~(\ref{A_Hamiltonian})] below.

When a finite hopping amplitude $t>0$ is switched on, the particles 
become delocalized from a single site (with finite or infinite 
localization length $\xi$). The hopping induces quantum fluctuations
of the occupation numbers $n_i$, and the on-site energies are renormalized 
by self-energy corrections of leading relative order $O[(t/\Delta)^2]$.
However, since the average occupation number on each 
site is essentially determined by the minimization of the local electrostatic
energy, the atomic limit approximation will still capture the essential
{\it static} physics of the \HA model (disorder screening and 
Mott-Hubbard physics), as long as $t\ll\Delta$, for arbitrary $U$.\cite{oytm08} 
In particular, this remains valid even for arbitrary localization length
$\xi$ (Ref.~\onlinecite{footnote0}) 
since the charge density may vary on the scale of a lattice
spacing, independently of the size of $\xi$. 
With these assumptions the \HA model is mapped 
onto an effective single-particle Anderson disorder model, 
\begin{eqnarray}\label{A_Hamiltonian}
 H & = & \sum\limits_{i\sigma} \!\left(\varepsilon_{i}^{\phantom{\dagger}}-\mu\right) c_{i\sigma}^{\dagger} c_{i\sigma}^{\phantom{\dagger}}  -  t\sum\limits_{<i,j>\sigma} c_{i\sigma}^{\dagger} c_{j\sigma}^{\phantom{\dagger}},
\end{eqnarray}
with the renormalized on-site energy distribution 
$p_{A}(\varepsilon)$.

In $d=1,2$ dimensions, as well as in $d=3$ dimensions for sufficiently strong
disorder, all particles described by Hamiltonian (\ref{A_Hamiltonian}) are exponentially localized.\cite{aalr79} The decay of their
wave functions $\psi(r)$, in the limit $r\rightarrow\infty$, is governed by
the localization length $\xi$. To analyze the effect of the repulsive
interaction on the localization, we study the $U$ dependence of $\xi$. For 
a first qualitative estimate we used in Ref.~\onlinecite{hks08} 
the relation\cite{tho72} (from now on we choose units where $a=t=1$)
\begin{eqnarray}\label{eqn thouless relation}
 \xi_1^{-1} & = & \int\limits_{-\infty}^{\infty} N(\varepsilon) \, \log\left|E-\varepsilon\right| \: d\varepsilon \nonumber \\
 & \approx & \int\limits_{-\infty}^{\infty} p_{A}(\varepsilon) \, \log\left|E-\varepsilon+\mu\right| \, d\varepsilon ,
\end{eqnarray}
valid in one dimension, 
where $\xi_1$ is the wave-function decay length,
$N(\varepsilon)$ denotes the disorder-averaged DOS, and $E$ is the particle energy measured relatively to the chemical potential.
The second approximate equality in Eq.~(\ref{eqn thouless relation}) holds for $\Delta \gg t$. 
For sufficiently large disorder, Eq.~(\ref{eqn thouless relation}) 
becomes also a good approximation in
$d>1$ dimensions.\cite{mk92} In Ref.~[\onlinecite{hks08}] we found that for all lattice fillings $\rho$
the localization length of a particle at the Fermi level, $\xi_{1}$, 
is a nonmonotonic function of $U$ with a pronounced maximum at intermediate $U$.

\section{Self-consistent Transport Theory}\label{sec sc}
In order to extend our analysis of the effective single-particle system [Eq.~(\ref{A_Hamiltonian})] to two- and three-dimensional systems as well as to parameter regimes with large localization lengths and to the calculation of general transport properties, one must go beyond the restrictions of one dimension and strong disorder
implied by Eq.~(\ref{eqn thouless relation}). Therefore, we study the system within the
self-consistent theory of Anderson localization.\cite{vw80}$^{-}$\cite{kop84}
This theory constitutes a resummation of the most divergent (Cooperon) contributions
to the irreducible particle-hole vertex, leading to a self-consistent equation
for the dynamical diffusion coefficient. The theory was originally 
developed for weak disorder and was extended to arbitrary 
disorder later on.\cite{kop84,kkw90,kro90} By comparison with direct numerical 
diagonalization results,\cite{bsk87} it was demonstrated that 
for the non-interacting Anderson model this theory 
yields quantitatively correct results for the phase diagram of Anderson
localization in $d=3$ and for the localization length in $d=1,~2,~3$
dimensions (with exception of the critical regime). 

For the interacting case with short-range, instantaneous 
interaction $U$, it is believed on general grounds that Fermi-liquid 
theory remains valid in the presence of disorder. Then the diffusion 
pole structure of the density propagator is preserved for particles at the Fermi energy ($E=0$) for temperature $T=0$. This was recently shown with the use of Ward identities\cite{kns08} to hold strictly at least when disorder- and interaction-induced self-energy contributions may be taken to be additive.   

In the formulation of Refs.~\onlinecite{kkw90} and \onlinecite{kro90} the self-consistent equation for the diffusion coefficient $D(\omega,E)$ reads
\begin{eqnarray}\label{sc eqn diff coefficient}
  & & D(\omega,E) \: = \: D_{0}(E) + \frac{2{\rm Im}\Sigma(E)}{[{\rm Im}G(E)]^2 D_{0}(E)} 
 \int\frac{d^{d}k}{(2\pi)^{d}} \nonumber \\
  & & \times  
\int\frac{d^{d}k'}{(2\pi)^{d}}  (\mathbf{v}_{\mathbf{k}}\cdot\hat{\mathbf{q}})
  \frac{{\rm Im}G_{\mathbf{k}}(E)\: [{\rm Im} G_{\mathbf{k}'}(E)]^2}
        {(\mathbf{k}+\mathbf{k}')^2_{p} - i\omega/D(\omega,E)}  
        (\mathbf{v}_{\mathbf{k'}}\cdot\hat{\mathbf{q}}), \nonumber \\
\end{eqnarray}
where $D_{0}(E)$ is the bare diffusion constant, 
\begin{eqnarray}\label{bare diff constant}
  D_{0}(E) & = & -\frac{1}{{\rm Im}G(E)} \int\frac{d^{d}k}{(2\pi)^{d}} 
(\mathbf{v}_{\mathbf{k}}\cdot\hat{\mathbf{q}})^2 [{\rm Im}G_{\mathbf{k}}(E)]^2
\end{eqnarray}
and $\hat{q}$ is the unit vector in the direction of the transport. The
disorder-averaged retarded single-particle propagators are given in terms of the self-energy $\Sigma(E)$ as
\begin{equation} 
G_{\mathbf{k}}(E) = [E + \mu - \varepsilon_{\mathbf{k}}  - \Sigma(E)]^{-1},
\label{green}
\end{equation}
and
$G(E) = \int\frac{d^{d}k}{(2\pi)^{d}} G_{\mathbf{k}}(E)$. 
$\varepsilon_{k} = -2\sum_{i=1}^{d} \cos(k_{i})$ is the dispersion and
$\mathbf{v}_{\mathbf{k}} = \nabla_{\mathbf{k}}\varepsilon_{\mathbf{k}}$ is the group velocity. 

In Eq.~(\ref{sc eqn diff coefficient}) the diffusion pole structure of the
integral kernel holds strictly for $Q=|\mathbf{k}+\mathbf{k}'| \ll 2\pi /\ell$,
where $\ell$ is the elastic mean-free path. For larger $Q$ the integral kernel
does not vanish but behaves in a non-singular way. Therefore, the  
momentum integrals in Eq.~(\ref{sc eqn diff coefficient}) must not be cut off
for $Q>2\pi/\ell$ but extend over the complete first Brillouin zone of the 
lattice.\cite{kkw90,kro90} 
Furthermore, one has to keep in mind that not only the single-particle
Green's function but also the particle-hole propagator obey 
lattice periodicity with respect to their momentum arguments.
In particular, the particle-hole propagator, which enters into the 
integral kernel of Eq.~(\ref{sc eqn diff coefficient}),\cite{vw92,kro90} 
is lattice periodic 
with respect to the center-of-mass momentum of the particle-hole pair. 
However, in Eq.~(\ref{sc eqn diff coefficient}) this 
periodicity is not explicit since the diffusion pole form of the kernel
arises from a hydrodynamic
expansion for small $Q$ and $\omega$. To restore the lattice periodicity
in the transport properties,
the subscript $p$ in the denominator of the kernel of 
Eq.~(\ref{sc eqn diff coefficient}) implies a 
shift by a reciprocal-lattice vector so as to keep the 
momentum argument $\mathbf{k}+\mathbf{k}'$ within the first Brillouin zone.

The localization length $\xi$ for particles at the Fermi energy is defined as the exponential decay length of the density correlation function in the static limit,\cite{vw92, footnote1} 
\begin{eqnarray}\label{eqn sc loc length}
\xi & = & \lim_{\omega\rightarrow 0} \sqrt{\frac{D(\omega,0)}{-i\omega}}, 
\end{eqnarray}
where in the localized phase the diffusion coefficient is purely imaginary 
to first order in $\omega$ and vanishes linearly for $\omega\rightarrow 0$.

For the evaluation of the theory, we first calculate the
disorder-averaged single-particle quantities, i.e., the 
self-energy $\Sigma(E)$ and the local Green's function $G(E)$, 
within the well-known coherent-potential approximation\cite{ym73} (CPA).
The CPA is known to interpolate disorder-averaged single-particle quantities correctly 
between the limits of weak and strong disorders, neglecting only exponentially
rare disorder configurations (Lifshitz tails of the DOS), and provides
quantitatively reliable results for disorder-averaged single-particle quantities over the
complete parameter range.\cite{kro90}
It is defined in connection with Eq.~(\ref{green}) and the renormalized 
level distribution $p_{A}(\varepsilon)$ by the self-consistent
relation,
\begin{eqnarray}\label{eqn cpa}
  \int d\varepsilon \, p_{A}(\varepsilon) \frac{\varepsilon - \Sigma(E)}{1 - [\varepsilon - \Sigma(E)]G(E)} = 0.
\end{eqnarray}

After the single-particle quantities are determined, the diffusion coefficient
$D(\omega\to 0,0)$ or the localization length $\xi $, respectively, is calculated by solving numerically  Eqs.~(\ref{sc eqn diff coefficient}) and (\ref{bare diff constant}) with Eq.~(\ref{eqn sc loc length}). The two static interaction effects, disorder screening and Mott-Hubbard gap formation, are incorporated in the single-particle transport theory through the quantities, $G_{\mathbf{k}}(E=0)$ and $\Sigma (E=0)$, determined by the renormalized distribution $p_{A}(\varepsilon)$. 

In our static treatment of the interaction term, the hopping of a particle at the Fermi level was assumed to happen on a background of immobile particles. In the case of singly occupied sites, the spin of these particles was considered to be randomly distributed. Therefore, in the absence of the site-energy disorder, i.e., $\Delta\rightarrow 0$, Eq.~(\ref{eqn cpa}) reduces to the Hubbard III approximation\cite{hub64,vke68} and, consequently, can be understood as an average over both spin and site-energy disorder. Implications and restrictions on the applicability of our approach will be mentioned in the discussion of our results in the following sections.

In $d>1$, the evaluation of the 2D integrals in 
Eq.~(\ref{sc eqn diff coefficient}) is numerically costly. 
Taking advantage of the periodicity of the integrands, 
this problem can be overcome by rewriting the factors in the 
integrand as Fourier series and using the convolution theorem. 
While for weak disorder the evaluation of the Fourier series is 
numerically still somewhat costly because of the pronounced peak structure of the Green's functions, the Fourier series converges quickly for larger disorder. The solution of the self-consistent equation [Eq.~(\ref{sc eqn diff coefficient})] is easily performed on a single desktop computer.

\section{Results and Discussion}\label{sec results}

The transport theory described above allows one to analyze in detail the 
static disorder screening effect of the Hubbard repulsion $U$ in the
\HA model in arbitrary dimension. 

\subsection {Two-dimensional systems}
Figure~\ref{fig dos} shows the 2D DOS $N(E)$, 
computed within CPA from Eq. (\ref{eqn cpa}), 
for a fixed disorder strength $\Delta$ and different values of the 
repulsion $U$. 
The figure clearly exhibits the regime of interaction-induced 
screening of disorder for small values of $U$, 
characterized by a narrowing of the disorder-averaged DOS with increasing $U$
[Figs. \ref{fig dos}(a) and \ref{fig dos}(b)]. It also shows the regime of hopping suppression 
for large $U$, where the Mott-Hubbard gap gradually develops with 
increasing $U$ [Figs. \ref{fig dos}(b)--\ref{fig dos}(d)]. The crossover between the two
regimes occurs roughly at $U\approx \Delta$. Note that the
present static approximation does not describe the Kondo-like 
quasiparticle resonance at the Fermi level
that would be induced by dynamical processes in high dimensions
near half filling.\cite{gkkr96} Therefore, it does not capture the typical many-body effects at the metal-insulator transition and inside the correlated metallic phase, known to be important when $\Delta\rightarrow 0$. 

\begin{figure}
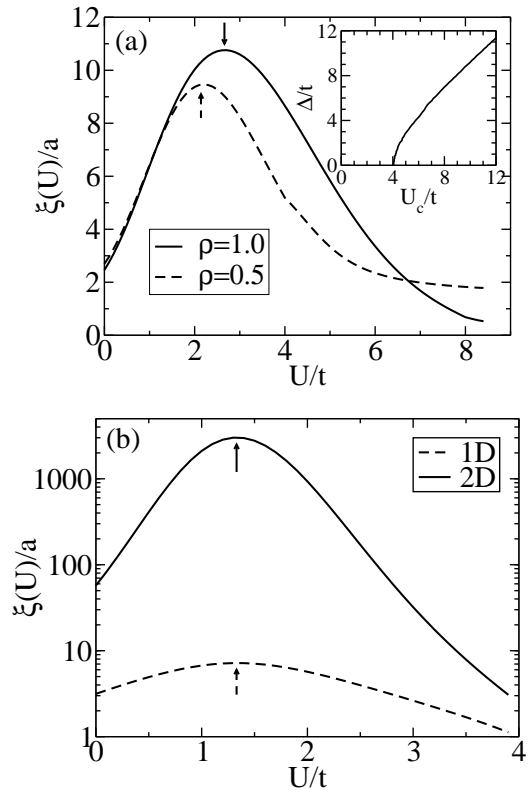

 \begin{center}
  {\includegraphics[scale=0.28,clip]{./hks_figure02a.eps}  \\~\\

  {\includegraphics[scale=0.28,clip]{./hks_figure02b.eps} }}
  \caption{
(a) Localization length $\xi$ at the Fermi level $(E=0)$ in $d=2$  
  as a function of $U$ for $\Delta/t=8$ and two different band fillings. 
  The arrows indicate the positions of the maximum of $\xi$ as obtained 
  from Eq.~(\ref{eqn thouless relation}) for a 1D system (Ref.~\onlinecite{hks08}).
  At the Mott transition for $\rho =1$ 
  [vanishing $N(0)$ due to Mott-Hubbard gap formation], $\xi(U)$ vanishes 
  exponentially as $\ln\xi(U) \simeq - 1/2N(0)$  because of the logarithmic 
  divergence of the Cooperon integral in Eq.~(\ref{sc eqn diff coefficient}) in 
  $d=2$. The inset shows the critical interaction 
strength, $U_c(\Delta)$, for the Mott transition at half filling. 
(b) Localization length $\xi$ at the Fermi level as a function of $U$ in $d=1$ and $2$ at half filling, $\rho =1$, for $\Delta/t=4$. The exponential enhancement of $\xi_{{\rm 2D}} (U)\sim \exp [1/\Delta_{\rm eff}^2]$ in $d=2$ due to 
  interactions is clearly seen while in 
  $d=1$ the dependence is comparatively weak (Refs.~\onlinecite{vw92} and \onlinecite{kro90}),
  $\xi_{{\rm 1D}}\sim [\Delta_{\rm eff}(U)]^{-1/2}$ (see text).
} 
\label{fig xiU}
 \end{center}
\end{figure} 

Figure \ref{fig xiU} shows the generic behavior of the localization length
at the Fermi level, $\xi(U)$, as a function of the repulsion $U$ for (a) 
strong and (b) intermediate disorders $\Delta$. 
The most salient feature seen in Fig.~\ref{fig xiU} is the nonmonotonic 
behavior of the localization length with a pronounced maximum at an
intermediate value $0\leq U_{\xi}\leq\Delta$. Within the one-dimensional (1D)
or strong disorder\cite{mk92} 
approximation [Eq.~(\ref{eqn thouless relation})],
$U_{\xi}$ can be calculated as $U_{\xi}\approx U_{\xi}^{({\rm
    1D})}=\Delta[\sqrt{1+3\rho(2-\rho)}-1]/3$.\cite{hks08} As seen in
Fig.~\ref{fig xiU}, this provides even in $d=2$ an excellent quantitative 
estimate for the results of the self-consistent theory not only for large 
disorder, as expected, but also for intermediate disorder. 
The nonmonotonic behavior is reproduced by numerical methods for 
finite-size systems, i.e., by quantum Monte Carlo (QMC) 
simulations\cite{sbs03,cds07} and statistical dynamical mean field 
theory (DMFT),\cite{swa08} with the maximum of $\xi(U)$ occurring almost 
precisely at $U_{\xi}^{(1D)}$. In the case of the QMC 
results\cite{cds07} the nonmonotonicity can be inferred 
from the dependence of the finite-temperature conductivity on $U$. In Ref.~[\onlinecite{swa08}] the on-site energies were 
calculated as poles of the atomic limit Green's function of the
\HA Hamiltonian, resulting in precisely the same Hamiltonian as our 
effective model [Eqs.~(\ref{eqn al distr}) and (\ref{A_Hamiltonian})]. 
In Ref.~\onlinecite{swa08} this Hamiltonian was then diagonalized 
numerically exactly for finite-size systems according to the statistical DMFT approach, and the 
localization length was extracted from the disorder-averaged inverse participation ratio 
(IPR) using the definition, 
\begin{eqnarray} 
\xi  & := & ({\rm IPR})^{-1/d} \ .
\label{eqn IPR}
\end{eqnarray}
We find good agreement with the results of Ref.~\onlinecite{swa08} for all
parameter values available up to a factor of order unity. This factor might be attributed to the slight difference
in their and our definitions of $\xi$ [Eqs.~(\ref{eqn sc loc length}) 
and (\ref{eqn IPR})], respectively. This agreement lends additional 
support to the quantitative correctness of the results of the 
self-consistent transport theory within the atomic limit approximation.

In our semianalytic theory the nonmonotonic  
behavior of $\xi$ is easily traced back to the
two competing interaction-induced effects: screening of the random potential
and DOS suppression due to a Mott-Hubbard gap.
Both effects are already incorporated in the
effective distribution $p_{A}(\varepsilon)$, as seen in Fig.~\ref{fig dos}: under an increase in $U$, the effective disorder is initially reduced, leading to an increase in $\xi$. For large $U$, however,
the formation of a Mott-Hubbard gap implies a reduction in the DOS at
the Fermi level and a broadening of the disorder distribution and, hence, a reduction in $\xi$ with increasing $U$ [Fig. \ref{fig xiU}(a)].

\begin{figure}
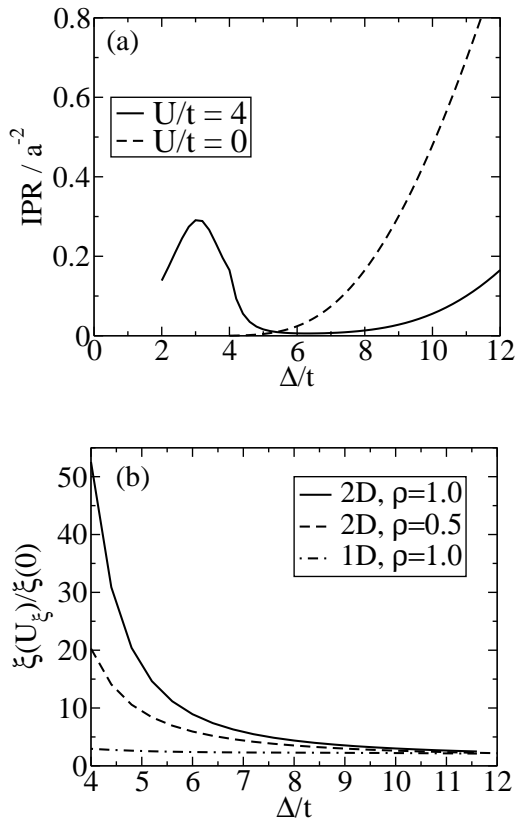

 \begin{center}
  \includegraphics[scale=0.28, clip]{./hks_figure03a.eps}
  \\~\\~\\
  \includegraphics[scale=0.28,clip]{./hks_figure03b.eps}
  \caption{
(a) Inverse participation ratio $\xi^{-2}$ as a function of the disorder
    strength $\Delta$ in $d=2$ for $U/t=4$ at half filling ($\rho=1$).
    The exponential suppression of the IPR for small disorder and the
    absence of the nonmonotonic behavior in the non-interacting case
    ($U=0$) are also shown (dashed line).
(b) Magnitude of the delocalization effect as a function of disorder strength
    in $d=1$ and $2$.}
\label{fig xiDelta}
 \end{center}
\end{figure} 

In Ref.~\onlinecite{ht04} the interaction strength was fixed and the IPR as a function of the disorder strength was studied. That work was based on a local unrestricted Hartree-Fock treatment
of the interaction and subsequent exact diagonalization of the resulting
effective disordered single-particle Hamiltonian. Our results show a nonmonotonic behavior 
of the IPR with a pronounced suppression (i.e., increase in $\xi$) 
for intermediate disorder, as seen in Fig.~\ref{fig xiDelta}(a), in good
agreement with the finite-size data of 
Ref.~\onlinecite{ht04}.\cite{footnote2}
Our theory indicates that this results from an intricate 
competition of Anderson localization and correlation-induced
disorder screening effects: for small increasing
$\Delta$ the IPR increases from zero due to Anderson localization in 
$d=2$ for arbitrarily weak disorder. As $\Delta$ increases further
to intermediate values, the effective disorder potential, derived from Eq.~(\ref{eqn al distr}),
is {\it reduced} by the screening effect until $\Delta \approx U$, 
resulting in a suppression of the IPR. Finally, when $\Delta > U$, 
the on-site repulsion $U$ cannot induce a further significant change in
local occupation numbers as $\Delta$ grows; hence the effective disorder potential and
the IPR increase.

Because of the logarithmic divergence of the 
Cooperon integral in Eq.~(\ref{sc eqn diff coefficient}) for $d=2$,
the disorder screening effect can induce an exponentially large effect
as seen in Fig.~\ref{fig xiU}(b). To demonstrate this 
screening-induced delocalization effect quantitatively, 
we show in Fig.~\ref{fig xiDelta}(b)
the ratio $\xi (U_{\xi})/\xi(U=0)$ of the maximal interaction-enhanced localization length and its non-interacting value.
The effect is especially strong for weak disorder in $d=2$, where 
$\xi$ depends exponentially on the screening-reduced 
disorder strength $\Delta_{\rm eff}$ of the effective Anderson model
[Eqs.~(\ref{eqn al distr}) and (\ref{A_Hamiltonian})] $\xi \sim \exp[1/\Delta_{\rm eff}^2]$.\cite{lr85} 

In Refs.~\onlinecite{ht04} and \onlinecite{cds07} also the observation of a metallic phase was reported. The existence of such a phase is not possible in an effective single-particle model\cite{aalr79} and would consequently be beyond our approach.  However, the good agreement of our results with the finite-size data of these works and the observation of the exponentially enlarged localization length, which exceeds the largest system size used in the numerics, suggests strongly that the infinite-size extrapolations erroneously indicated a true metallic state in $d=2$.\cite{fw08}

The remarkable quantitative agreement of our theory with the results of numerical calculations wherever comparison is possible suggests
that the essential physics of localization in the 2D \HA model is 
captured by the assumptions of static disorder screening and 
Mott-Hubbard gap formation.

\begin{figure}
 \begin{center}
  \includegraphics[scale=0.28]{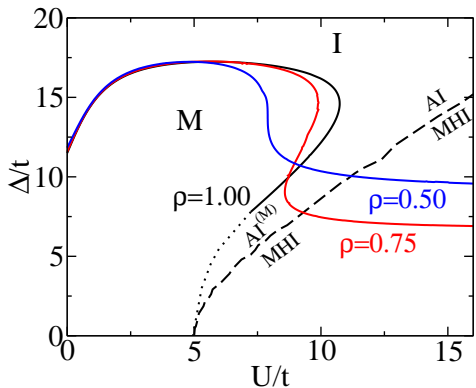}
  \caption{(Color online) 3D phase diagram in the $(U,\Delta)$ plane for 
different lattice fillings $\rho$. The metallic phase (M) is to the lower left while
the insulating phase (I) is to the upper right of the solid mobility edge
curves for $\rho=1.00$, $0.75$, and $0.50$, respectively. 
The dashed curve shows the critical interaction strength for the Mott transition at half filling.
The dotted line is an extrapolation of the
mobility edge for $\rho=1$ toward small disorder, where the evaluation of the 
self-consistent Eq.~(\ref{sc eqn diff coefficient}) becomes numerically 
costly. The corresponding Anderson, Mott-Hubbard, and Mott-Hubbard assisted Anderson insulating regions at half filling are marked by AI, MHI, and AI$^{\rm (M)}$, respectively.}\label{fig phasediagram}
 \end{center}
\end{figure}

\subsection{Three-dimensional systems}
In $d>2$, the non-interacting Anderson model describes a disorder induced
metal-insulator transition, where extended and localized states are separated
by the mobility edge.\cite{aalr79} Hence, the question about a possible
disorder reduction by interaction gets even more relevant. 
We have calculated the phase diagram of localization in $d=3$ using the
atomic limit approximation [Eqs.~(\ref{eqn al distr}) and (\ref{A_Hamiltonian})]
with the self-consistent transport theory of Anderson localization
[Eqs.~(\ref{sc eqn diff coefficient})--(\ref{eqn cpa}); see
Fig.~\ref{fig phasediagram}]. 

The figure shows, for different values of the lattice filling $\rho$, the disorder screening effect: above the critical disorder value for
Anderson localization without interaction
\cite{footnote3} $\Delta _c/t \approx 11.7$,
the increase in interaction leads to a re-entrance from the 
Anderson insulating into the metallic phase, i.e., the metallic phase 
is extended to larger values of $\Delta$. 
This behavior was also observed in a recent  DMFT
study\cite{bhv05} of the \HA model where a site-dependent self-energy
correction was used. Contrarily, generalized DMFT studies based on
site-independent averaged self-energy
corrections do not describe the screening effect.\cite{kns08} 

For the half-filled case ($\rho=1$) a further increase in 
$U$ eventually leads to a suppression of the averaged DOS at the
Fermi level, $N(0)$, because of the gradual formation of a Mott-Hubbard gap (dashed line). Consequently the system undergoes first a transition from the metallic phase to a (Mott-Hubbard assisted) Anderson insulator with a reduced but finite $N(0)$ just before the DOS vanishes at the dashed line and the Mott-Hubbard insulating phase is entered.
This intermediate phase is not likely to be 
seen in DMFT studies of the problem\cite{bhv05,kns08} because within DMFT  
the formation of a Kondo resonance and the concatenated unitarity sum 
rule for $N(0)$ in the metallic state prevent a Mott-Hubbard-induced 
suppression of $N(0)$ and lead to a complete screening of the disorder 
potential.\cite{bhv05,kns08} It remains to be seen whether in the physical
3D disordered systems Kondo physics or pseudogap formation 
dominates. 

Away from half filling, a pure Mott-Hubbard
transition is not possible because $N(0)$ remains non-zero. 
Therefore, for weak disorder the metallic phase extends out to 
arbitrarily large $U$. However, for large $U$ ($U/t\gtrsim 10$) 
increasing disorder $\Delta$ does induce an Anderson metal-insulator
transition, and this transition occurs at a substantially smaller
value of $\Delta$ than for small and intermediate $U/t\lesssim 7$
because of a Mott-Hubbard reduced DOS at the Fermi level (pseudogap).
It is interesting to note that this interplay of Anderson localization
and Mott-Hubbard physics away from half filling leads to a 
step-like behavior of the phase boundary, as seen in 
Fig.~\ref{fig phasediagram}, connecting smoothly to the phase boundary
of the half-filled case. 

\section{Conclusion}
We have presented a self-consistent study of the static disorder 
screening effect
and of Mott-Hubbard gap formation, induced by the local repulsion in the 
\HA model. Both of these effects are represented by an 
interaction-induced renormalization of the effective distribution of 
on-site energy levels and a subsequent mapping of the \HA model onto 
a single-particle Anderson model. While this mapping is exact in the atomic 
limit, we have argued that it still provides a good description of the
static screening effect for finite hopping amplitude even when
the localization length $\xi$ is large. The localization properties of the
effective single-particle Anderson model were then treated by the 
self-consistent theory of Anderson localization.\cite{vw80,kro90}
We found rich behavior of the localization length in two dimensions
and of the phase diagram in three dimensions due to an intricate 
interplay of disorder screening and Mott-Hubbard physics in the different regions of parameter space. 

Despite the technical simplicity of our approach, it yields good agreement with numerical studies\cite{sbs03}$^{-}$\cite{swa08} of the same problem for two-dimensional 
finite-size systems, including the non-monotonic dependence of $\xi$ 
on both the interaction strength $U$ 
and the disorder $\Delta$. At the same time, we found in $d=2$ an
exponential interaction-induced enhancement of $\xi$
for weak and intermediate values of $\Delta$ although a true metallic
state is not possible within our effective single-particle theory. 
The good agreement of our results with the numerical finite-size
calculations, and simultaneously, our prediction of a large but finite 
localization length show that the indications of a true metallic state,
found in some works by infinite-size extrapolations of the numerical data, are not conclusive.
These indications might rather be due to the fact that the largest
system sizes were still smaller than the localization length of the
infinite system. 

For three-dimensional systems we found that for weak Hubbard interaction
$U$ the disorder screening effect results in a re-entrant 
behavior from the insulating to the  metallic phase while for large $U$,
disorder and Hubbard interaction cooperate to form a 
Mott-Hubbard assisted Anderson insulating phase (with finite 
density of states at the Fermi level), which exists in a 
finite range between the metallic phase and the Mott-Hubbard-gapped phase
present for half filling.

\acknowledgments
This work was supported in part by the Deutsche Forschungsgemeinschaft through SFB 608.


\begin{thebibliography}{99}
\bibitem{footnote0}
Note that in $d=3$ $\xi$ diverges at the Anderson transition, 
and in $d=2$ it can become large even for $\Delta /t \sim O(10) \gg 1$, 
as seen in Sec. IV.
\bibitem{footnote3}
This value is somewhat smaller than the one obtained in 
Ref.~\onlinecite{kro90}, because there the lattice periodicity of the
particle-hole propagator in the center-of-mass momentum $Q$ was not 
considered.
\end{thebibliography}
\end{document}